\documentclass[preprint,journal]{vgtc}       

\ifpdf
  \pdfoutput=1\relax                   
  \usepackage{graphicx}                
  \DeclareGraphicsExtensions{.pdf,.png,.jpg,.jpeg} 
\else
  \usepackage{graphicx}                
  \DeclareGraphicsExtensions{.eps}     
\fi%

\graphicspath{{figures/}{pictures/}{images/}{./}} 

\usepackage{microtype}                 
\PassOptionsToPackage{warn}{textcomp}  
\usepackage{textcomp}                  
\usepackage{mathptmx}                  
\usepackage{times}                     
\usepackage{cite}                      
\usepackage{tabu}                      
\usepackage{booktabs}                  

\usepackage{algpseudocode}
\usepackage{algorithm}

\usepackage{enumitem}
\usepackage{wrapfig}
\usepackage[hidelinks, bookmarks=false]{hyperref}
\usepackage{xcolor}

\newcommand{\name}[1]{\texttt{DendroMap}}
\newcommand{\baseline}[1]{\texttt{t-SNE-Grid}}
\newcommand{\letterFigure}[2]{\textbf{(#1) #2}}

\preprinttext{}

\onlineid{0}

\title{DendroMap: Visual Exploration of Large-Scale Image Datasets\\for Machine Learning with Treemaps}

\author{Donald Bertucci, 
        Md Montaser Hamid,
        Yashwanthi Anand,
        Anita Ruangrotsakun,\\
        Delyar Tabatabai,
        Melissa Perez,
        and Minsuk Kahng}
\authorfooter{
\item
 The authors are with Oregon State University.
 E-mail: \{bertuccd,
          hamidmd,
          anandy,
          ruangroc,
          tabatase,
          peremeli,
          minsuk.kahng\}@oregonstate.edu.
\vspace{3pt} 
}

\shortauthortitle{Bertucci \MakeLowercase{\textit{et al.}}: DendroMap: Visual Exploration of Large-Scale Image Datasets for Machine Learning with Treemaps}

\abstract{
In this paper, we present \name{}, a novel approach to interactively exploring large-scale image datasets for machine learning (ML). ML practitioners often explore image datasets by generating a grid of images or projecting high-dimensional representations of images into 2-D using dimensionality reduction techniques (e.g., t-SNE). However, neither approach effectively scales to large datasets because images are ineffectively organized and interactions are insufficiently supported.
To address these challenges, we develop \name{} by adapting Treemaps, a well-known visualization technique. \name{} effectively organizes images by extracting hierarchical cluster structures from high-dimensional representations of images. It enables users to make sense of the overall distributions of datasets and interactively zoom into specific areas of interests at multiple levels of abstraction. 
Our case studies with widely-used image datasets for deep learning demonstrate that users can discover insights about datasets and trained models by examining the diversity of images, identifying underperforming subgroups, and analyzing classification errors.
We conducted a user study that evaluates the effectiveness of \name{} in grouping and searching tasks by comparing it with a gridified version of t-SNE and found that participants preferred \name{}. 
\name{} is available at \url{https://div-lab.github.io/dendromap/}.
} 


\teaser{
\centering
\vspace{12pt} 
\includegraphics[width=1.0\linewidth]{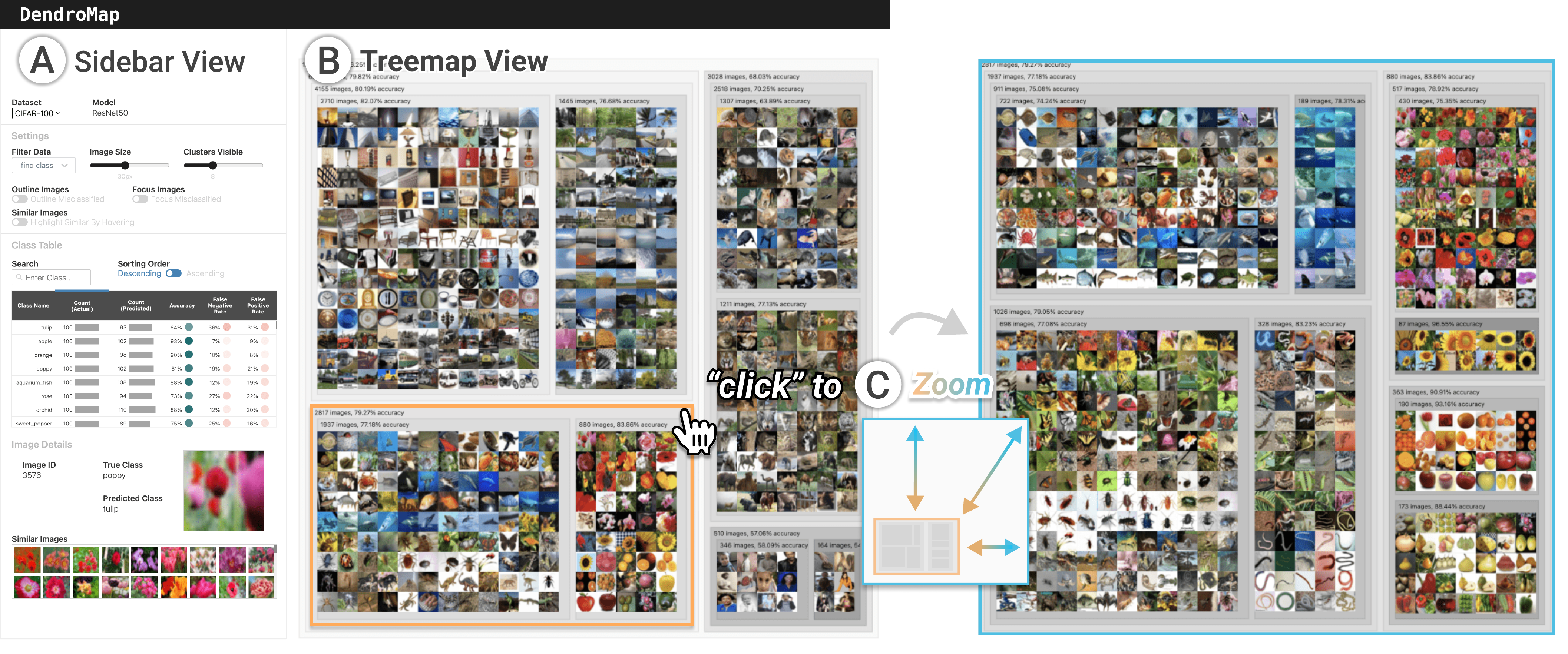}
\vspace{-20pt}
\caption{With \name{}, users can explore large-scale image datasets by overviewing the overall distributions and zooming down into hierarchies of image groups at multiple levels of abstraction.
In this example, we visualize images of the CIFAR-100 dataset by hierarchically clustering the image representations obtained from a ResNet50 image classification model. 
\letterFigure{B}{Treemap View} displays these clusters of images
organized as a hierarchical structure 
by adapting Treemaps.
By clicking on a cluster, a user can interactively \letterFigure{C}{Zoom} into that image group, revealing subgroups that replace and fill the available space with animation. 
The user clicked on a cluster for organism images, which creates
distinct subgroups of fish, insects, fruits, and flowers.
With \letterFigure{A}{Sidebar View}, the user can dynamically adjust the number of clusters to be displayed and inspect the class-level statistics.}
\label{fig:teaser}
\vspace{5pt} 
}

\renewcommand{\manuscriptnotetxt}{This paper has been accepted for the IEEE VIS 2022 Conference and will be published in the IEEE Transactions on Visualization and Computer Graphics.}

\nocopyrightspace

\vgtcinsertpkg

\begin{document}


\firstsection{Introduction}

\maketitle

\begin{figure*}
    \centering
    \includegraphics[width=0.9\linewidth]{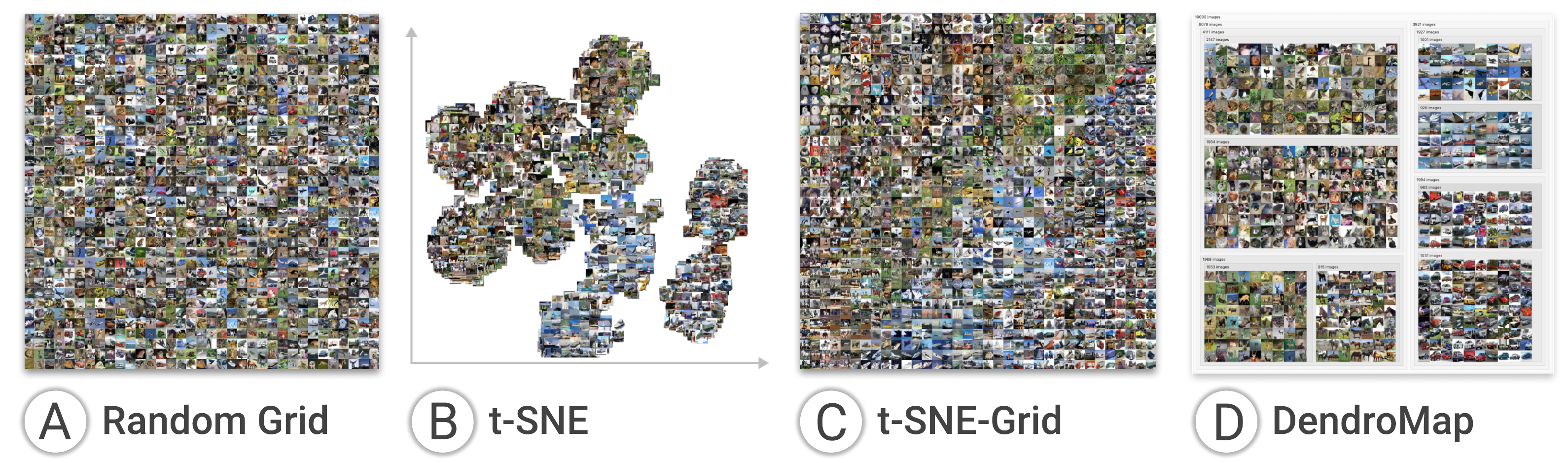}
    \vspace{-10pt}
    \caption{
    Commonly used approaches to visualizing image datasets for ML include 
    generating a grid of images (A);
    projecting the images onto 2-D displays using techniques like t-SNE (B); and 
    using a combination of these two approaches (C). 
However, they do not scale well to large datasets because images are ineffectively organized and interactions are insufficiently supported.
We present DendroMap (D) which effectively organizes image datasets using a modified interactive treemap algorithm. 
    }
    \label{fig:summary}
\end{figure*}

The machine learning (ML) community is increasingly aware of the importance of understanding datasets.
There is a growing interest in \textit{Data-Centric AI}, as opposed to the model-centric approach~\cite{ng2021chat,polyzotis2021can,whang2021data}.
A deep understanding of the datasets can help inform design decisions for building ML models efficiently and appropriately.
It motivates important decisions such as collecting more data, changing data labeling policy, debiasing models, and more. 

While images are used extensively in deep learning,
fundamental challenges exist in exploring image datasets because they lack attributes like those found in tabular data.
A commonly-used approach is to use dimensionality reduction (DR) techniques like t-SNE~\cite{maaten2008tsne} over multivariate features extracted from the datasets~\cite{rauber2016visualizing}.
To enable users to easily see the contents of the images, 
each data point in the projected space is often replaced with its corresponding image (like in \autoref{fig:summary}B)~\cite{smilkov2016embedding}. 
However, for large datasets, there could be overlaps between images, and inefficient use of space (i.e., a lot of white space) makes the size of each image too small for users to inspect.

To scale up DR methods for large image datasets, 
data points can be re-positioned into
a \textit{grid} such that no images overlap and the grid fills the entire screen 
(\autoref{fig:summary}C)~\cite{karpathy2014tsnegrid,tsnegrid,hilasaca2019distance}. It is inspired by a common approach to visualizing image collections---displaying images as a grid~\cite{zahalka2014towards,bederson2001photomesa}. This is also the case for ML image datasets. Many tutorials for image classification begin by displaying a sample of images as a grid~\cite{kerastutorial}.
While the combination of the grid and t-SNE methods effectively use 2-D space, 
it is still severely limited by the size of the image datasets. 
Adding interactions may help, but the use of \textit{semantic zooming} is not straightforward for the gridified version of t-SNE.
This is because optimization algorithms 
were applied that distort the original space~\cite{jv1987jv,tsnegrid}, which means the before-zoom and after-zoom versions may present very different sets of images.

In this paper, we present \name{}, a novel interactive visualization for exploring large-scale image datasets by adapting treemaps, a well-known visualization technique. \name{} effectively organizes images using hierarchical clustering algorithms and displays the hierarchical structure with an interactive treemap. 
A set of image clusters provides an overview of a dataset, and users can interactively zoom into the clusters to investigate sub-clusters in the hierarchy. 
\autoref{fig:teaser} illustrates an example.
It initially displays eight clusters, each showing a sample of images whose size is proportional to the total number of images in that cluster.
Unlike traditional treemaps, the number of image clusters showing can be dynamically changed by the user to customize the level of abstraction. Furthermore, clicking on a cluster will zoom (\autoref{fig:teaser}C) into that cluster to reveal and fill the space with sub-clusters.

\name{} aims to support a wide range of analytics tasks for ML practitioners.
This includes bias and error analysis at the instance and subgroup levels, which have been identified as important in the literature~\cite{hohman2018visual,amershi2015modeltracker,cabrera2019fairvis}.
To build highly accurate and less biased models, it is crucial to have datasets containing a diverse set of images.
\name{} enable users to categorize the types of images present in the datasets and estimate their distributions.
Furthermore, \name{} users can identify underperforming subgroups for error analysis~\cite{wu2019errudite,olson2021contrastive,zahalka2020ii}.

\vspace{3pt}
\noindent\textbf{Contributions}. 
The contributions of this paper are as follows:
\begin{itemize}[itemsep=2pt,topsep=0pt,parsep=0pt]
    \item \name{}, \textbf{a novel interactive visualization system for exploring large-scale
    image datasets used in ML.} \name{}
    adapts an interactive zoomable treemap and supports the information seeking mantra ``overview first, zoom and filter, then details-on-demand''~\cite{shneiderman2003eyes}. Combined with the sidebar as a part of multiple coordinated views, ML pracitioners can perform a wide range of tasks for data-centric analysis (e.g., error analysis, bias discovery).

    \item \textbf{An adapted treemap algorithm for hierarchical dendrogram
    structures of images,} which allows users to dynamically specify the number of clusters to visualize, enabling exploration at multiple levels of granularity.
    Images are systematically sampled to fill the space for each cluster, providing an overview of the datasets. 
    
    \item \textbf{Live demo on the web}\footnote{The live demo of \name{} is available at \url{https://div-lab.github.io/dendromap/}.} with available code\footnote{The code is available at \url{https://github.com/div-lab/dendromap}.} and \textbf{use cases} for \name{} demonstrating users' dataset exploration, bias discovery, and error analyses.
    
    \item \textbf{A quantitative user study} designed to 
    compare \name{} with a gridified version of t-SNE, a space-filling technique used by ML practitioners. 
    Participants performed a wide range of grouping tasks and preferred \name{} over the baseline.
    
\end{itemize}

\section{Related Work}
\label{sec:related}

\subsection{Visualization for Machine Learning}

Visualization has helped ML practitioners perform a variety of analytics tasks such as: exploring datasets, analyzing performance results, interpreting and explaining model internals, building models, monitoring training progress, and debugging models~\cite{hohman2018visual, yuan2021survey}.

Many existing visualization tools for ML support the tasks of analyzing performance results and exploring datasets at multiple levels of abstraction, ranging from individual instances to entire classes.
While ML practitioners often only use summary metrics (e.g., accuracy) or class-level statistics, visualization researchers have argued the importance of instance-level analysis.
Early works include ModelTracker~\cite{amershi2015modeltracker}, Squares~\cite{ren2016squares}, and Facets-Dive~\cite{facets, wexler2019if}.
These tools represent each instance as a small square using the \textit{unit visualization} technique~\cite{park2017atom}, enabling users to see individual instances in the context of aggregated information.
This can work particularly well for image datasets as each square can be replaced with a thumbnail of the actual image content.

While instance-level analysis has benefits, 
the scale of datasets urges researchers to develop ways to slice and filter datasets, resulting in subgroup-level analysis~\cite{kahng2017activis, hohman2018visual,he2021can}.
This allows users to specify data subsets based on attributes and perform more fine-grained analysis than at the class-level. 
However, image data creates a fundamental challenge in supporting such analysis because there are no attributes beyond class labels.
Therefore, group structures are often created with algorithmic approaches. A common approach is to use a DR technique like t-SNE~\cite{maaten2008tsne} or UMAP~\cite{mcinnes2018umap}, which are often applied to high-dimensional representations obtained from neuron activations~\cite{rauber2016visualizing}. 
We propose an alternative approach to capturing group structures.

ML researchers have stressed the importance of datasets by coining terms like Data-Centric AI and MLOps~\cite{ng2021chat}.
Our work aligns with this trend to ensure that ML datasets are less biased, more fair and inclusive, and contain fewer errors.
A recently developed tool named Know Your Data~\cite{kyd} aligns with this goal,
providing statistics based on attributes obtained from external APIs (e.g., face recognition, object detection). Our work instead focuses on making sense of raw image datasets by relying on human perception.

\subsection{Image Browsing}
Zah\'alka and Worring~\cite{zahalka2014towards} presented a comprehensive overview of multimedia visualization methods (primarily of images) in their survey.
They categorized existing techniques into five types:
basic grid, 
similarity space, 
similarity-based, 
spreadsheet, 
and 
thread-based.
The three methods commonly used by ML practitioners described in Sect. 1 and \autoref{fig:summary} (i.e., random grid, t-SNE, and a grid version of t-SNE) belong to the ``basic grid,'' ``similarity space,'' and ``similarity-based'' categories, respectively.
Our proposed treemap-based method can also be placed in the ``similarity-based'' category.

The idea of using treemaps for image browsing was proposed in PhotoMesa~\cite{bederson2001photomesa}.
It consists of two variations of the treemap algorithms: the ordered treemap algorithm ensures the order of images in each treemap block will match the order in file structures (e.g., by timestamp); and the quantum treemap ensures that the widths and heights of the generated rectangles are integer multiples of a
given elemental size. 
Unlike the data commonly used in treemaps, ML datasets have different properties: 
each dataset has a set of classes, and the images within each class have no order.
Because there is no existing hierarchical structure, we extract one using agglomerative clustering algorithms.

An important task in analyzing images or multimedia data is categorizing or exploratory searching.
The key difference from tabular datasets is that image datasets are not annotated with structured attributes---images are unstructured.
Many common data operations like filtering, grouping, and sorting cannot be easily applied.
If we consider low-level tasks by Amar et al.~\cite{amar2005low}, only a few of the 10 tasks can be applied to images~\cite{zahalka2014towards}.
Thus, an important challenge in interactive visualization of image data is automatic extraction of semantic information, interactive exploration of categories, or both~\cite{van2016iclic,zahalka2020ii,xie2018semantic}.

\subsection{Similarity-based Visualization Methods}

As we discussed in the previous subsection, our proposed work can be considered as a similarity-based approach.
We briefly describe both the similarity-space and similarity-based approaches in the ML context.

The t-SNE algorithm is probably the most popular among ML researchers. It is often used to visualize cluster structures learned by deep learning models~\cite{maaten2008tsne,rauber2016visualizing,wang2020understanding,chatzimparmpas2020t}.
While t-SNE often plots each data point as a small circle in a 2-D space, 
the nature of images provides us with the opportunity to directly plot a small thumbnail instead of a dot.
This enables users to see the image contents without interacting with each circle mark (e.g., clicking, hovering).
For example, Embedding Projector~\cite{smilkov2016embedding} displays MNIST images in t-SNE plots.
However, as the number of images grows, images overlap, making it almost impossible to see them in high-density areas (see Fig.~\ref{fig:summary}B).

Researchers and practitioners have devised methods to address the issue of overlapping images.
The images can be  rearranged in a grid either by
selecting a sample of images among many in each grid or
redistributing all images into all the grid spaces in screen using optimization algorithms~\cite{jv1987jv}.
Although we have not found research papers to gridify t-SNE or UMAP, there exist several implementations~\cite{karpathy2014tsnegrid,tsnegrid,ml4atsne}, including one by Karpathy~\cite{karpathy2014tsnegrid}.
This type of gridifying algorithm has been used in several visual analytics tools for ML for image data~\cite{zhao2021human,chen2020oodanalyzer,wattenberg2016use}.

Redistributing data points or images into a rectangular grid has also been studied in non-ML context, such as  IsoMatch~\cite{fried2015isomatch} and rectangular packing~\cite{gomi2008cat}.
Removing overlaps can be more intelligent by balancing the full use of screen space and intentionally leaving some white-space to reveal cluster structures~\cite{hilasaca2019overlap}.

\subsection{Hierarchical Exploration of Data}

To begin our review of hierarchical exploration, we provide a brief background about clustering algorithms~\cite{murtagh1983survey}.
Unlike the $k$-means algorithm which 
partitions data points into a fixed number of groups,
the hierarchical clustering algorithms iteratively divide data space into smaller space (i.e., divisive) or merge from smaller groups into larger groups (i.e., agglomerative).
We use the latter to form a hierarchy (called a \textit{dendrogram}), since divisive does not produce high-quality results for high-dimensional data and is computationally expensive for large data.
The agglomerative ones align more closely with useful characteristics of t-SNE: focusing on similar pairs to find cluster structures.

Existing work on visualizing dendrograms include
Hierarchical Clustering Explorer (HCE) ~\cite{seo2002interactively},
Stacked Trees which interactively merge  parts of the dendrogram~\cite{bisson2012improving},
and Yang et al. for steering and revising the dendrograms~\cite{yang2020interactive}.
All these used node-link diagrams to display dendrograms; however, they are less desirable for image datasets, because the dendrograms require all instances to be positioned along a single line, which means the size of images would become very small if we want to display images in place of the leaves of the dendrogram tree.
A space-filling technique like treemaps can resolve this issue.

Hierarchical data exploration has been studied extensively in text domains.
Text data is unstructured, so automatic extraction of clusters is important too like images.
HierarchicalTopics~\cite{dou2013hierarchicaltopics} extracts  hierarchical structures of latent topics and enables users to explore and revise them.
TopicLens~\cite{kim2016topiclens} allows users to zoom into certain areas of projected two-dimensional spaces.
Marcilio et al. extracts hierarchical structures from high-dimensional representations of deep learning data~\cite{marcilio2021explorertree}. 
Nmap represents data as treemap-style representations, similar to ours~\cite{duarte2014nmap}. It adjusts initial positions of data items obtained from 2-D projection algorithms by iteratively creating treemap nodes using their modified slicing algorithm.
We instead create tree structures using well-known clustering algorithms. Another difference is that our work targeting image data displays image thumbnails within treemap nodes.

\begin{figure*}[t]
    \centering
    \includegraphics[width=0.85\linewidth]{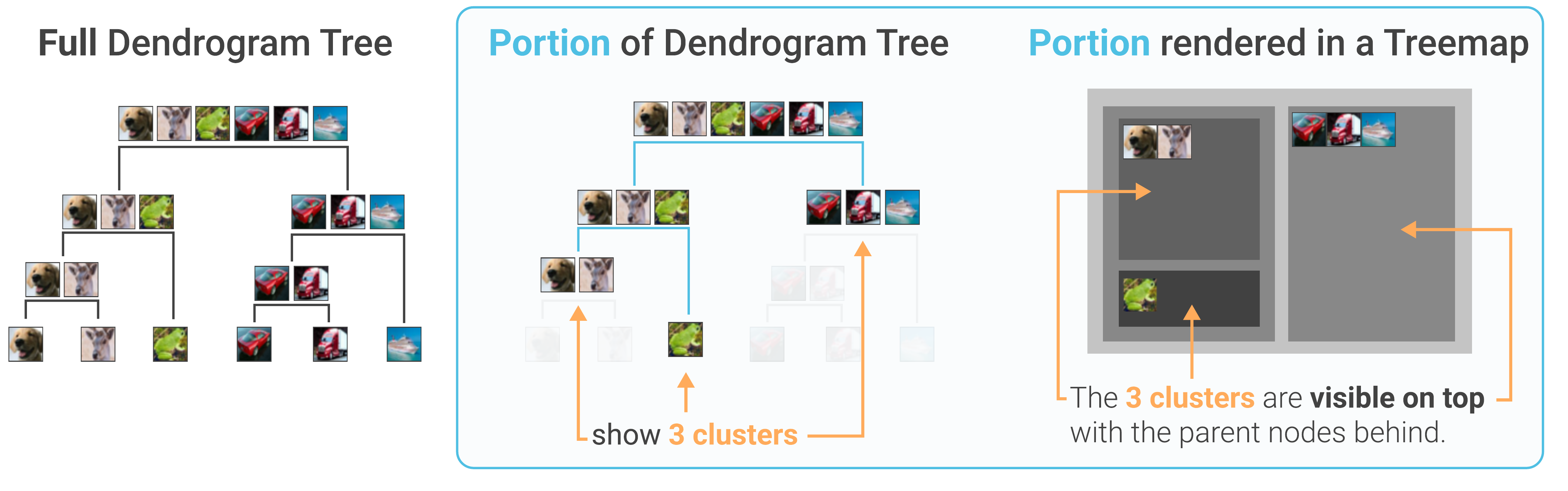}
    \vspace{-13pt}
    \caption{To scalably visualize the dendrogram tree structure created from agglomerative clustering methods, users can dynamically specify the number of clusters to be rendered  in \name{}. In this example, a portion of the dendrogram is rendered in the treemap view to show three image clusters. Increasing the number of clusters to be shown will result in creating more partitions across the treemap with smooth animations.}
    \vspace{-4pt}
    \label{fig:summaryTreemap}
\end{figure*}

\section{Design Goals}

To help ML practitioners explore large-scale image datasets, we adapt treemaps with the following design goals:

\begin{enumerate}[topsep=1pt,itemsep=1pt,parsep=0pt]
\item
\textbf{Overview of Data Distributions.} 
We aim to assist users in getting an overview of datasets as a beginning step for their analysis of datasets. 
This includes helping them answer questions like what kinds of images mostly exist in their datasets, whether they are \textit{diverse} enough~\cite{hong2020crowdsourcing} or biased towards any properties~\cite{buolamwini2018gender}. 

\item 
\textbf{Exploring at Multiple Levels of Abstraction.}
We aim to design our visualization to provide users with abilities to interactively adjust the level of abstraction.
While treemaps are effective at supporting \textit{abstract and elaborate} interactions~\cite{yi2007toward},
we adapt the original treemap techniques by considering unique properties of the dendrogram structure and the domain of ML for images.

\item \textbf{Instance-level Exploration.} As images do not contain attributes, it is important for users to see the individual image contents while exploring datasets. 
We aim to effectively organize image thumbnails to help users find and inspect individual data points while they navigate over the tree structure.

\item \textbf{Subgroup-level Analysis for ML.}
Both the literature in multimedia analytics and visual analytics for ML point out the importance of identifying subgroups from datasets~\cite{zahalka2014towards,hohman2018visual,olson2021contrastive}.
This can be useful for performing a wide range of analytic tasks in ML, such as error analysis and bias discovery~\cite{wu2019errudite,cabrera2019fairvis}.

\end{enumerate}

\section{\name{} Construction and Interactions}

This section describes how a dendrogram can be constructed from an image dataset, how \name{} visualizes the dendrogram, and how supported interactions help achieve our design goals.

\subsection{Dendrogram Tree Construction}
\label{sec:dendrogram}
To create groups of images for hierarchical exploration, we use the well-known hierarchical agglomerative clustering algorithm~\cite{murtagh1983survey}.

The clustering algorithm takes as input high-dimensional representations of images.
There are several ways to obtain such representations, such as by extracting high-dimensional embeddings from pre-trained or fine-tuned models, low-dimensional encodings using Autoencoders, or raw image pixels~\cite{rauber2016visualizing, hohman2018visual, kahng2017activis, hinton2006reducing}.
In our user study in \autoref{sec:user-study}, for the CIFAR-10 dataset, we extracted 1024 dimensional embedding vector representations from the second-to-last hidden (fully-connected) layer in pretrained ResNet50 models that we fine-tuned on CIFAR-10. 

Given this input,
each image vector is initialized as its own cluster to start, then the most similar image clusters are merged together using Ward linkage with the Euclidean distance metric to form more balanced trees~\cite{murtagh1983survey}. The agglomerative merging process repeats until the final two clusters merge into one cluster containing all the images in the dataset.
The output of the algorithm forms a special tree structure, called \textit{dendrogram}, with leaf nodes corresponding to data instances.

\subsection{\name{} Visualization}
\label{sec:treemap}

\name{} visualizes dendrogram structures using a modified treemap algorithm. It traverses the dendrogram and renders each cluster node as a grid of images using the available rectangular space. 

\textbf{Treemap Layout}.
The dendrogram resembles a binary tree, and all non-leaf nodes have only two child nodes. This allows \name{} to adapt the traditional slice-dice treemap layout~\cite{shneiderman1992treemap}. Normally, slice-dice creates undesirable aspect ratios when laying out many rectangles per level~\cite{bederson2002ordered}; however, this issue does not occur in ours because the dendrogram will not have more than two children per node, always resulting in just one partition of space.

\begin{figure}[!tb]
    \centering
    \includegraphics[width=0.75\linewidth]{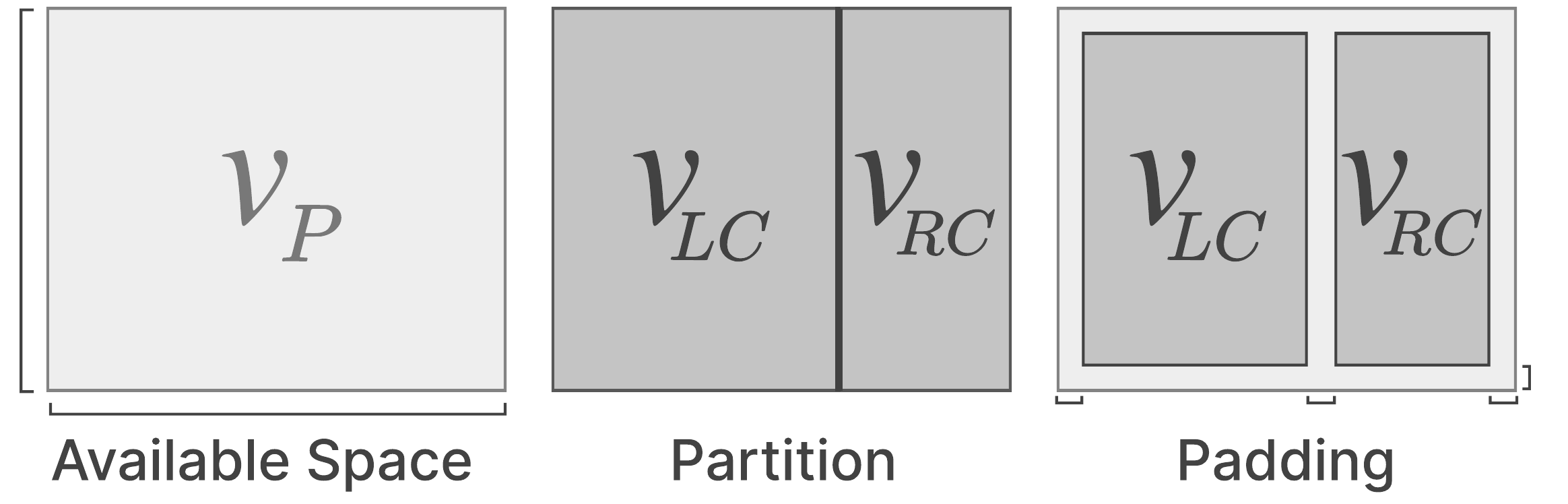}
    \vspace{-10pt}
    \caption{The slice-dice layout takes the available space given by the parent node $v_P$ and partitions the space into for its two children $v_{LC}$ and $v_{RC}$. To reveal the $v_P$'s hierarchy, padding is added to the children boxes.
    }
    \vspace{-3pt}
    \label{fig:layout}
\end{figure}

We modify the slice-dice layout to display a grid of fixed sized images on top and to include padding (to highlight hierarchical structures). To demonstrate one iteration of the modified layout, consider a node $v_P$ that has two children $v_{LC}$ and $v_{RC}$ with 6 and 4 images, respectively. The goal is to fill a 100 by 90 pixel available space depicted in \autoref{fig:layout}. The algorithm works as follows:
\begin{enumerate}[topsep=1pt,itemsep=0pt,parsep=0pt]
    \item \textbf{Dice if the available space from the parent $v_P$ is a horizontal rectangle and slice if it is vertical.} In \autoref{fig:layout}, $v_P$'s width $w_P$ is 100 pixels and height $h_P$ is 90 pixels, so dicing is chosen.
    \item \textbf{Compute the ratio to partition the space}. When dicing, the partition ratio is calculated by $ratio := N_{LC} / N_{P}$, where $N_i$ represents the number of images in $v_i$. The left and right areas of the partition correspond to each child, $v_{LC}$ and $v_{RC}$. In \autoref{fig:layout}, the dice partition ratio is computed as $(6 / 10) = 0.6$. Meaning $60\%$ of the space is for the $v_{LC}$ and $40\%$ is for $v_{RC}$.
    \item \textbf{Adjust the partition to fit images}.
    Based on the image size, compute the maximum amount of the images that can fit across entire parent's width (or height if slicing) by $fit := \lfloor w_{P} / w_{image}  \rfloor$, where $w_P$ is the width of the available space for $v_P$ and $w_{image}$ is the width of each image. Then the actual partition dimensions can be calculated as $\lfloor fit \times ratio \rfloor$ pixels, resulting in a partition that fits images without cutting them off.
    \item \textbf{Add padding to show hierarchies}. After laying out the $v_{LC}$ and $v_{RC}$ and assigning them their new dimensions, a fixed padding is added to reveal the parent cluster $v_P$ behind it (like in \autoref{fig:layout}). 
    We set 
    a fixed padding of 10 pixels in our implementation. 
    Color can encode
    the remaining height of tree under that node
    \cite{bostock2019nestedtreemap}.
\end{enumerate}

\textbf{Adjusting the number of clusters}.
Traversing the \textit{entire} dendrogram quickly fills the available screen space, making it hard to display many images. Thanks to the dendrogram's binary tree structure, each iteration of the \name{} algorithm only lays out two children (one partition), which allows us to render specific number of clusters (i.e., $k$ set by users).
By traversing the tree breadth-first and counting the $k$ clusters created so far, the algorithm can stop and show those $k$ clusters.
For example in \autoref{fig:summaryTreemap} the dendrogram traversal stops to only render three clusters showing in the treemap. 

\textbf{Organizing images within the clusters}.
A useful property of dendrograms is that the leaf nodes (i.e., images) are positioned along a line based on the structure of the constructed tree in a way that there is no edge crossing.
We use this positional information to organize the list of images for each cluster node.
As seen in the \autoref{fig:summaryTreemap} dendrogram, the similar images merge together starting from the bottom, and at each successive merge, they still maintain the position of the leaves in the dendrogram from left to right. The end result is the root node cluster's images are in the same position as the leaf nodes in the dendrogram,
which lets similar looking images clump up together and nearby images in a cluster be likely more similar than images located far within the cluster.
For example, in \autoref{fig:teaser} on the right, insect images taken over white background are clustered together with a large node.
When there exist a larger number of images to display than the amount of available space,
we systematically sample images from the cluster. 
Specifically, we compute the period by calculating the total number of images in the cluster over the maximum number of images we can possibly show and round down to the nearest integer. We then sample images in the cluster with that period of frequency. For example, if $30$ images can be shown in a given space and the cluster has $150$ images, we compute the period to sample by $\lfloor 150 / 30 \rfloor = 5$ and iterate over the cluster $\{ x_1, ..., x_{150} \}$. The end result is an image every $5$ iteration to determine $30$ images $\{x_1, x_6, x_{11}, ..., x_{146}\}$.
This enables us to show representative samples of a cluster and avoid hiding images that occur later on.
We display 
the total number of images at the top of each cluster node, as well as their classification accuracy if available.

\textbf{Zooming interactions.}
\label{sec:zoom}
For further navigation of the clusters, \name{} supports zooming. 
When a cluster node is clicked, \name{} animates 
to zoom into the new cluster, which enlarges the selected cluster to fit into the entire space, and creates a set of subclusters within the selected cluster.
Our implementation basically follows Bostock's zoomable treemap implementation~\cite{bostock2019zoomabletreemap}. 
In addition, by taking up the entire space with the zoom-in, more images can be shown with more specific hierarchies, leading to more in-depth exploration. This process corresponds to rendering a downstream portion of the dendrogram. At any point, by clicking back on the parent cluster, the reverse process of zooming-out goes back up the tree to reveal the top-level view again.
These zooming interactions allow users to quickly explore large image collections at multiple levels of granularity.

\subsection{Coordinated Views with the Sidebar}
\label{sec:sidebar}

We developed a system for \name{} by designing coordinated views consisting of the main treemap view and the sidebar.
The sidebar contains rendering settings for the treemap display, a class table for class-level error analysis, and a panel for details for a selected image.

\textbf{\name{} Settings}.
The sidebar contains two sliders to change the overview level: one controls the number of clusters visible and the other controls the image size. 
By default, \name{} shows eight clusters of medium-sized images to balance the level of detail and overview such that many images can be shown while still separated into distinguishable groups.
These sliders allow users to easily change the overview level based on their exploration needs. 

When a dataset comes with predictions from a trained model, 
the sidebar provides two options to highlight misclassified images. One toggle highlights these images using a red border and the other toggle puts the images into focus by making the others translucent. 
Visually emphasizing misclassified images makes it easier for users to find groups of images that the model consistently misclassifies.

\textbf{Class Table.}
The class table is visible if model predictions are present. The table contains information for additional \textit{error analysis} at the class level. The table updates based on the parent cluster's images (i.e., the root or previously selected cluster; by default, all images). 
Each row of the table corresponds to a specific class in the dataset (e.g., cat). 
The next two columns of the table displays the counts of images with a true or predicted class label matching the class specified.
\begin{figure}[!t]
    \centering
    \includegraphics[width=\linewidth]{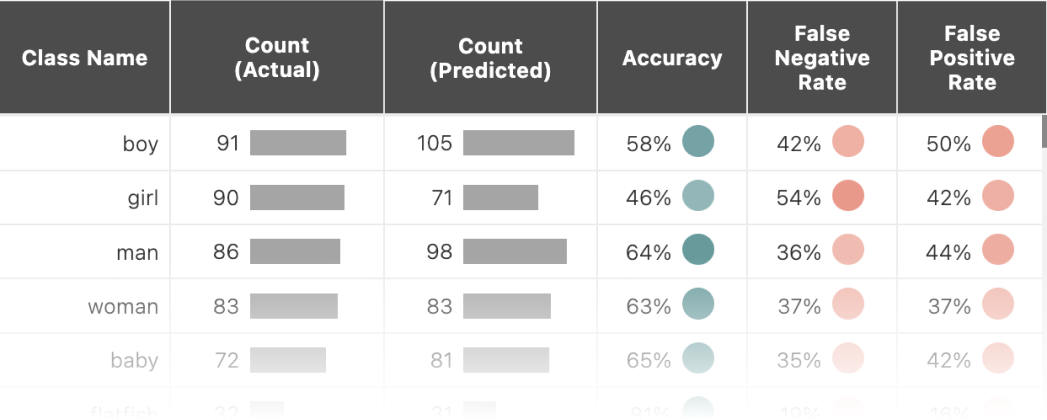}
    \vspace{-19pt}
    \caption{The class table summarizes class-level statistics of images present in the selected cluster in the treemap view.
    The user can sort and search for classes, and hover over each entry to quickly locate accurate or error filled clusters highlighted directly on the \name{}.}
    \vspace{-4pt}
    \label{fig:classtable}
\end{figure}

The last three columns of the table provide useful metrics for class-level error analysis: the prediction accuracy (i.e., how often the true and predicted classes matched that row's class), the false negative rate (i.e, how often the true class matched that row's class but the predicted class was different), and the false positive rate (i.e., how often the predicted class matched that row's class but the true class was different). As shown in \autoref{fig:classtable}, each rate is encoded with the opacity of a colored dot.

By hovering over one of these entries in the table, the treemap view highlights the images used to determine that metric by making the other images translucent. This way users can use the class table in tandem with the treemap to isolate and find areas of high error or high accuracy.

\textbf{Image Details.}
A user can click on an image in \name{} to see detailed information: larger view of the image, true class label, predicted class label if it has one, and similar images.
The similar images are determined based on distances in the high-dimensional space, which can be used for counterfactual analysis~\cite{cheng2020dece,gomez2020vice}.

\subsection{Implementation Details}

The \name{} system was built using  \textit{Svelte}\footnote{Svelte JavaScript Framework: \url{https://svelte.dev/}}, a reactive JavaScript framework that has been increasingly used in the visualization community.
The main component, the treemap view, is implemented primarily with \textit{D3.js}\footnote{D3 JavaScript Library: \url{https://d3js.org/}} to create SVG elements and to transition the elements for natural animation. 
The complimentary component, the sidebar, is entirely implemented in \textit{Svelte}, and uses \textit{Svelte} \texttt{store}  functionality to communicate between the treemap. 
The dendrogram structure is created from the SciPy\footnote{SciPy Python Library: \url{https://scipy.org}} hierarchical clustering implementation with Ward linkage (recommended as default). The output dendrogram is exported as a nested JSON object to be rendered as a treemap on the client side. 

\section{Use Cases}

In this section, we describe how \name{} can be used in practice to explore and analyze image datasets through three usage scenarios.

\begin{figure*}[t]
    \centering
    \includegraphics[width=0.85\linewidth]{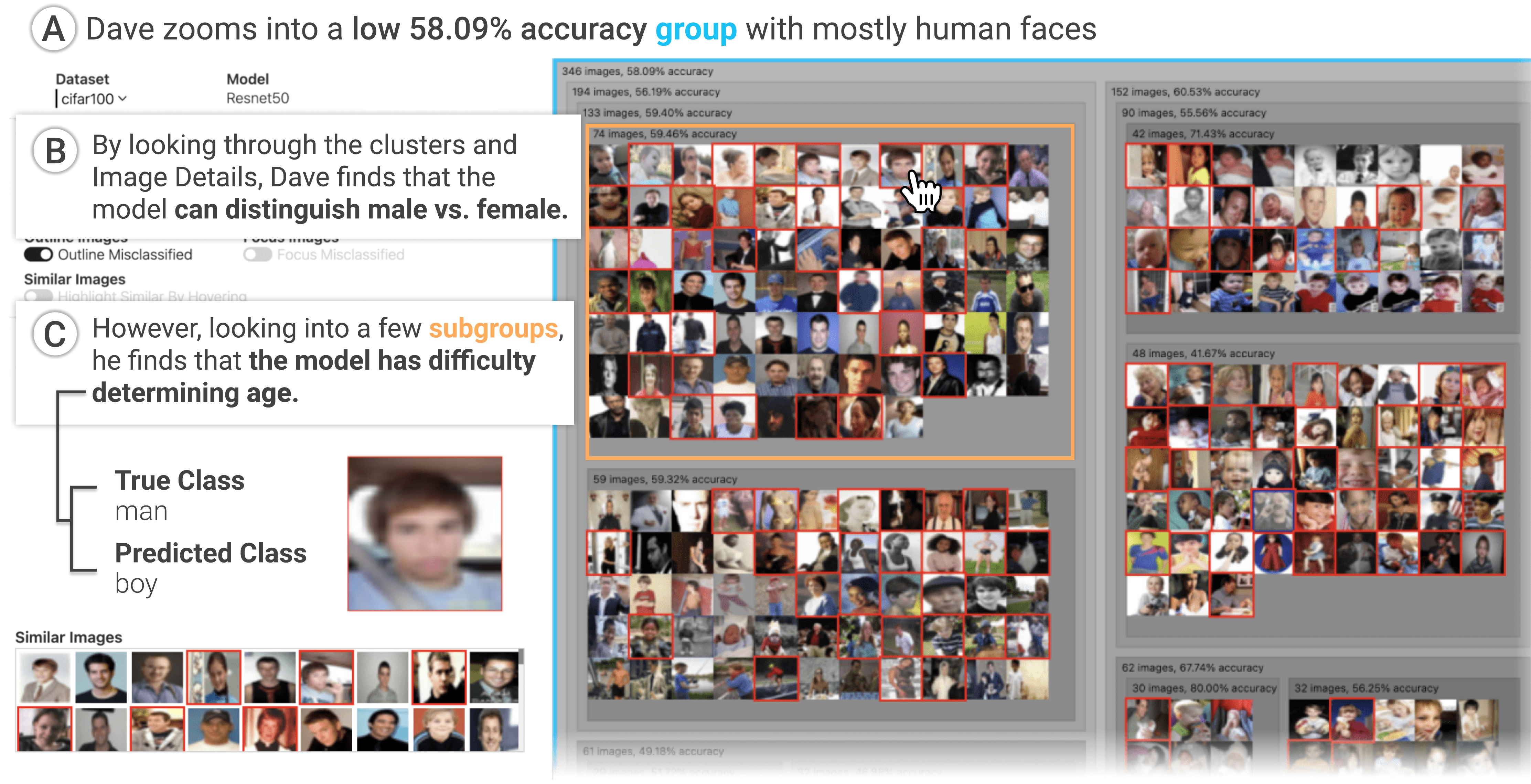}
    \vspace{-10pt}
    \caption{In our case study, ML practitioner, Dave, 
    investigates the specific classes that his model struggles with using \name{}. 
    }
    \label{fig:scenario}
\end{figure*}

\subsection{Examining Bias in Datasets}

Consider Priya, a data scientist who lives in the Southeast region of Asia and is evaluating whether ImageNet can be used to train an image classification model that she can deploy in her country.
After she loads the \name{} interface, Priya begins to click around to ``zoom'' into different portions of the dataset. 
She first clicks on the rectangle containing the approximately half of the dataset and discovers a cluster containing everyday objects. 
She notices a cluster of taxi cabs and hovers over the class name ``taxicab'' in the sidebar's class table to put just the taxicab photos in focus while the rest become faded. 
She notices that most are black or yellow, but she knows from personal experience that many taxis are multicolored in her country, so she makes a note to supplement the ``taxicab'' class with some of those images. 
Priya ``zooms out'' by clicking on the outermost rectangle and decides to visit another cluster, this one featuring many images of people interacting with a variety of everyday objects, such as ``violin'' and ``sunscreen''. 
However, as she clicks on several images to get a better look at each one, she notices that the images tend to include people with lighter skin tones.
She makes another note to supplement the dataset with images of people with darker skin tones interacting with the objects corresponding to each of the classes listed in the class table.
Given these notes, Priya now would like to train models using this dataset and evaluate them by a set of slices which she made notes (e.g., skin tone), to make sure the models perform consistently over these slices.

\subsection{Identifying Underperforming Subgroups}

Consider Dave, a ML engineer who is using the CIFAR-100 dataset to evaluate a trained image classification model.
He opens \name{} and sees the default view of eight rectangles or clusters. 
As Dave inspects the interface, he notices that the group of images with the lowest accuracy score (57 percent) consists mostly of human faces.
He sees no obvious pattern at this level of overview in the hierarchical structure, so he clicks on another rectangle to get a closer look. 
From the class table in the sidebar, he observes that a majority of the images in this group were predicted to be ``woman'' or ``girl'', but most were incorrect. 
Dave thinks perhaps his classification model has trouble determining which of those two labels is correct.
He navigates back up one level by clicking on the outermost rectangle. 
He selects a different cluster and this time he observes that a majority of the images are predicted as ``man'' or ``boy'', but with similar proportions of incorrect guesses (as shown in \autoref{fig:scenario}).
From these two insights, Dave hypothesizes that his model can distinguish male and female faces, but has difficulty determining whether the person is a child or adult. 
Then he decides to collect additional training data of human faces for four different categories: adult female, adult male, boys, and girls.

\subsection{Analyzing Classification Errors}

Consider Anna, a ML practitioner who 
works in a team developing computer vision applications.
While she trained a model, she noticed her model consistently had a harder time correctly predicting images from the artifact-related classes
so she decided to analyze her model for these classes
from the ImageNet dataset, such as ``umbrella'' and ``frying pan''.
She opens \name{} and toggles the ``outline misclassified'' and ``focus misclassified'' switches to spotlight the misclassified images, outlined in red, while the others fade.
She notices that the red outlined images appear to be scattered without much of a pattern, so she gradually increases the number of clusters until \name{} splits the images into subgroups of higher or lower accuracy. 
She stops when it reaches 18 clusters because she notices distinct subgroups of images with high accuracy (over 90 percent). 
Most of these subgroups focus on particular classes, such as ``racket'' or ``potter's wheel''.
Anna wants to investigate the cause of clusters with much lower prediction accuracy, so she continually clicks on the next visible cluster with the lowest accuracy.
She notices a pattern as she keeps drilling down towards the leaf nodes: the accuracy rate decreases as the images become more cluttered.
She clicks on several misclassified images to inspect their true  and predicted class labels, and she discovers that the predicted labels are not necessarily inaccurate---it is that the true label and predicted labels are classifying the entire image based on only a portion of it.
For example, she clicks on an image of a couple of people sitting on a bench on a sunny day. The true class label for this image is ``sunglasses'' because one person is wearing sunglasses, whereas the predicted label for the image is ``park bench'' because the two people are sitting on a bench.
These errors can be critical for her team's applications, so Anna decides to consider object detection models which can locate multiple objects within a single image, instead of image classification models.

\section{User Study}
\label{sec:user-study}

To evaluate the effectiveness of \name{} for exploring large-scale machine learning datasets, we conducted a user study comparing \name{} and a baseline visualization technique for images, \baseline{}, a gridified version of t-SNE.

\subsection{Baseline: \baseline{}}
\label{sec:tsnegrid}
We compare \name{} with a gridified version of t-SNE, which we call \baseline{}.
It re-adjusts the positions obtained from the t-SNE algorithm~\cite{maaten2008tsne}, by filling the available rectangular grid space with the images for effectively using screen space~\cite{karpathy2014tsnegrid}. 

\begin{figure}[!b]
    \centering
    \includegraphics[width=0.9\linewidth]{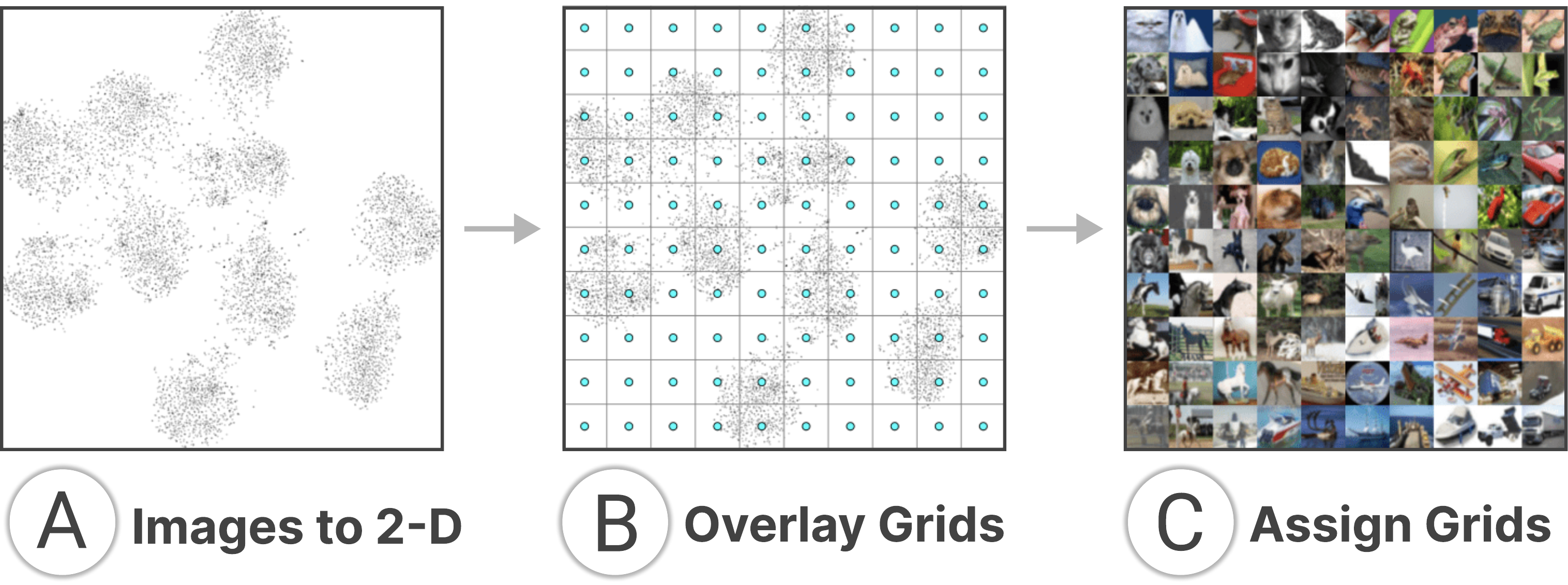}
    \vspace{-8px}
    \caption{Steps to generate \baseline{}: From t-SNE embeddings (in \textbf{A}),
    we first overlay grid points on top of the embeddings
(in \textbf{B};
$10 \times 10$ in this case). Then in \textbf{C}, 
we assign each grid with an image that has the smallest distance.
} \label{fig:summaryBaseline}
\end{figure}

This process works by first taking the image representations from the dataset and reducing them down to their two-dimensional embeddings using t-SNE (like Fig.~\ref{fig:summaryBaseline}A). Then, to fill the space, two dimensional grid points are evenly laid out over the space of image embeddings (like Fig.~\ref{fig:summaryBaseline}B). Finally, each grid point is assigned the closest image embedding and the corresponding image is displayed on top (like Fig.~\ref{fig:summaryBaseline}C). The result is a grid of images with the structure from t-SNE.

There may be overlap with what is considered the closest image embedding to each grid point, so to achieve a result where the sum of grid assignment distances is minimized, the Jonker-Volgenant algorithm is used to get the optimal assignments~\cite{jv1987jv}.
The optimal grid assignments work by phrasing the problem as a linear assignment problem. 
For this user study, to enhance the \baseline{} exploration further, we implemented a one-level zoom that recomputes the grid with a smaller number of images based on where the user clicks in \baseline{}. In particular, the top-$k$ closest to the click are recomputed with the Jonker-Volgenant algorithm to display a smaller and more focused grid of images to the user, where $k$ is chosen based on the number of grids to show in the zoomed-in view. For example, to show a $5 \times 5$ grid, $k$ is set as $25$ to take the $25$ closest points and gridify them. We open-sourced the grid assignment implementation and published it as a library\footnote{\url{https://www.npmjs.com/package/grid-assign-js}}.
\subsection{Study Setup}

\subsubsection{Participants.}
We recruited 20 participants by using the departmental student mailing lists.
Their average age was 26. 
Five were female and 15 were male. 
Six were undergraduate and 14 were graduate students. Their degree programs included computer science, robotics, and AI. 
We recruited only those who 
have taken at least one AI or ML course. 
Every participant attended the study in-person and we had one participant per session. 
Each participant was compensated with a \$20 gift card.

\subsubsection{Protocol.}
We used a within-subject design such that each participant  evaluated both \name{} and \baseline{}. 
To let participants work with different images for the two visualizations, we created two variations of the CIFAR-100 dataset (Artifact and Organism subsets which
we describe in detail in ~\autoref{sec:datasets-and-models}).
From the two visualizations and two datasets, we created four conditions. Each participant was assigned to one of these four conditions to ensure there was no bias in the order in which a participant used 
(shown in \autoref{table:conditions}). 

\begin{table}[!b]
\centering
\begin{tabular}{c|c|c|c|c}
\toprule
\# & \multicolumn{2}{c|}{Phase 1} & \multicolumn{2}{c}{Phase 2}\\\cmidrule{2-5}
 & Visualization & Dataset & Visualization & Dataset\\\midrule
1 & \baseline{} & Artifact & \name{} & Organism\\
2 & \name{} & Artifact & \baseline{} & Organism\\
3 & \baseline{} & Organism & \name{} & Artifact\\
4 & \name{} & Organism & \baseline{} & Artifact\\
\bottomrule
\end{tabular}
\vspace{2pt}
\caption{Four conditions for counterbalancing the orders of two interfaces in our within-subject design}
\label{table:conditions}
\end{table}

\begin{table}[!t]
\centering
\begin{tabular}{@{\hskip 3pt} c @{\hskip 5pt} p{3.1in}}
\toprule
\# & Task Description\\
\midrule
1. & \textbf{Categorizing images} into groups across 40 classes\\
2. & \textbf{Categorizing images} into groups for a single class\\
3. & \textbf{Identifying groups} of images with high classification accuracy within a single class\\
4. & Estimating the image count \textbf{distribution} over multiple groups within a single class\\
5. & \textbf{Searching} for an image with a given text description\\
6. & \textbf{Searching} for an image with a given visual description\\
7. & Searching for an \textbf{anomalous} image with an incorrect class label\\
\bottomrule
\end{tabular}
\vspace{3pt}
\caption{Seven tasks designed to evaluate several grouping and searching tasks used in ML analysis}
\label{table:tasks}
\end{table}

Every participant completed two sets of tasks, one for each visualization-dataset combination of their respective condition. 
For each phase, a participant was given a brief tutorial of the visualization, then they were asked to complete seven tasks while thinking aloud. 
After each phase, the participant filled out a post-questionnaire form. 
All participants used the same computer setup with a 32-inch monitor.

\subsubsection{Dataset and Models.}
\label{sec:datasets-and-models}
We used the CIFAR-10 and CIFAR-100 datasets \cite{krizhevsky2009learning} for the study. The CIFAR-10 dataset has 10 classes, each containing 6,000 images (5,000 from training set and 1,000 from test set), while the CIFAR-100 dataset has 100 classes, each containing 600 images. 

We fine-tuned the ResNet50 \cite{DBLP:conf/cvpr/HeZRS16} architecture that was pretrained on the ImageNet dataset provided by TensorFlow\footnote{\url{https://www.tensorflow.org/api_docs/python/tf/keras/applications/resnet50/ResNet50}}. The CIFAR-10 and CIFAR-100 images were upsampled to fit the input shape of the ResNet50 model (i.e., $224 \times 224 \times 3$). After extracting the image features from the models, we used Average Pooling, followed by three Dense layers (i.e., their sizes are 1024, 512, and the number of classes, respectively). The model was fine-tuned for 20 epochs, achieving a test set accuracy of $92.8 \%$ on CIFAR-10 and $76.3 \%$ on CIFAR-100. For use in the \name{} and \baseline{} algorithms, we represented the images in each dataset as high-dimensional vectors from embeddings of one of the last hidden layers in each respective model
(i.e., for CIFAR-10, the second-to-last hidden layer, which is  1024-dimensional; for CIFAR-100, the last hidden layer, which is 512-dimensional\footnote{For CIFAR-10, we chose the layer farther from the output layer, because we wanted to extract lower-level concepts that are less specific to classes for people to explore different types of images within each class~\cite{bau2017network}.}).
The \name{} and \baseline{} use the same representations as input to their respective algorithms.

We divided the classes of CIFAR-100 into two sets--``Artifacts'' and ``Organisms''--in order to have two distinct sets of classes for the within-subject design. This helps ensure that results from the first interface only minimally affect those from the second interface. Each set consists of 40 classes (i.e., 4 superclasses, each consisting of 10 classes)~\cite{krizhevsky2009learning}.
The Artifact set contains classes like chair, television, and bottles, while the Organisms set contains classes like tiger, crocodile, and trout.

\subsubsection{Tasks}
The participants completed seven tasks which can be divided into two broad categories: grouping and searching. The grouping tasks involved identifying groups of images based on semantically similar properties; the searching tasks involved searching for images based on specific properties. Table \ref{table:tasks} provides a summarized description of the tasks.

\begin{itemize}[itemsep=0pt,topsep=0pt,leftmargin=8pt,parsep=0pt]
\item In Tasks 1 and 2, participants were asked to categorize images into 3-4 groups 
based on semantically similar properties. Task 1 was designed to evaluate how users make sense of and categorize images across many (i.e., 40) classes whereas Task 2 focuses on how users make sense of images within a single class. The common objectives of these two tasks include analyzing diversity or any potential bias present in the distribution of the data as well as getting an overview of the data.

\item In Task 3, we asked participants to find two large groups, using images from a single class, that have very high classification accuracy and have specific properties. 
This task was designed to evaluate the scope of subgroup-level error analysis.

\item Task 4 is about examining the distribution of images for a single class. This task was designed based on the ``characterize distribution'' task discussed by Amar et al. \cite{amar2005low}.  The participants were asked to estimate the approximate proportions of four groups determined based on an attribute (e.g., color of objects).

\item The following two tasks 
are conventional searching tasks.
In Task 5, participants were asked to find an image that matches a provided text description.
In Task 6, participants were asked to find the image that matches the one on the task sheet.

\item Lastly, Task 7 was designed to find probable anomalies.
Participants were asked to find potential labeling errors among the misclassified images for a single class~\cite{northcutt2021pervasive,xiang2019interactive}. 
\end{itemize}
Note that every participant worked with the same task list for both \name{} and \baseline{}, but used a different dataset for each of the visualizations.

\subsubsection{Interface Setup}

For fairer comparison, the sidebar component from \name{} was added to the \baseline{} visualization. Additionally, to confirm that certain sidebar components are not overused over the main visualization, the class table, class filtering, and similar images components were removed from the sidebar for both \name{} and \baseline{}.

\subsection{Results}

The setup of our user study gives us the scope to analyze data from a multitude of perspectives.

\subsubsection{Evaluation of task completion time}
Our first set of analyses focused on task completion time. During the study, we recorded the time a participant took to complete each task. 
We conducted Wilcoxon signed-rank tests, and there is no significant difference between the average time taken by our participants with \baseline{} and that with \name{} for each of all the tasks.

\subsubsection{Evaluation of task responses}
We evaluated the responses to the seven tasks using statistical methods.

\textbf{Task 1.}
We instructed our participants to identify four groups such that an image can be assigned to only one group (\textit{mutually exclusive}) and most images
present in the interface can be assigned to one of the groups (\textit{collectively exhaustive}). To evaluate the quality of groups made by the participants, we conducted three analyses. 
First, to measure the collectively exhaustive property of the groups, we counted the number of classes covered by at least one of the four groups
and divided that number by the total number of classes present in the dataset (i.e., 40). 
The reason why we counted the number of ``classes'' instead of ``images'' is
the number of classes can approximate the number of images because each class has an equal number of images. 
In an ideal scenario, the value would be 1.0. 
If only a portion of images in a class belongs to a group, we count it as half.
With \name{}, the average value over all participants are higher with a value of $0.82$, compared to $0.73$ with \baseline{}. 
A one-sided Wilcoxon signed-rank test 
indicates that its $p$-value is 0.089.
This suggests that on average, participants were able to maintain the ``collectively exhaustive'' property more with \name{} than \baseline{}, but we note that the level of significance is not high.
Next, to assess the mutual exclusiveness of the groups made by a participant, we counted the number of classes that belong to two or more groups.
In an ideal scenario, the value is zero because there is no overlap between the groups. 
We calculated the average value to be $0.07$ for \baseline{} and $0.13$ for \name{}. The results of the same test
indicate that on average participants were able to create more ``mutually exclusive'' groups with \baseline{} than \name{} ($p$-value = 0.062). 
Lastly, we calculated the entropy score of the probability distribution of the four groups to check how much the groups are equally distributed. 
We found the average entropy score of \name{} to be similar to that of \baseline{} (i.e., $1.37$ vs. $1.34$). 

\textbf{Task 2.} Like Task 1, the participants were asked to identify mutually exclusive and collectively exhaustive groups. The main difference for Task 2 is that they worked with images for only one class.
To evaluate the quality of groups identified by the participants, we conducted the same three analyses as for Task 1. However, for Task 2, instead of counting the number of classes, we labeled a 10\% sample of individual images.
In our first analysis of the collectively exhaustive property, the average values for \baseline{}  and \name{} are almost the same with the values of $0.67$ and $0.66$ respectively. This also happened with the mutual exclusiveness analysis
(i.e., $0.10$ and $0.13$).
Our final analysis of the entropy scores is also no exception 
(i.e., $1.41$ 
and 
$1.36$). 

\textbf{Task 3.} This task is also about grouping, but the participants were asked to find two large groups of images with 
\textit{high classification accuracy}. 
We conducted two analyses. 
First, 
we assessed the average accuracy of the two groups. To find the accuracy of each group, we counted the correctly classified images from the total number of images covered by each group. The average accuracy values of the two groups are $92.2\%$ and $93.2\%$ for \baseline{} and \name{}, respectively. \name{} is slightly higher, but there is no significant difference.
Second, 
we measured the size of these groups.
The average for \baseline{} is $0.38$ and for \name{} is $0.34$, with no significant difference.

\textbf{Task 4.} In this task, the participants estimated the approximate percentage of different cars and birds based on 
    car color (yellow, red, white or silver, or other) or 
    background of birds (e.g., sky), respectively.
To evaluate their responses, we counted the number of car and bird images that correspond with 
the aforementioned criteria 
and calculated the Kullback-Leibler (KL) divergence score to quantify how much the probability distributions reported by our participants differ from the actual distributions. A score of 0 means the two distributions are the same. 
Our results show that \name{} has more counts in between 0.0 and 0.1 than \baseline{} (i.e., 10 vs. 7). This indicates that more participants were closer to the actual distribution when using \name{}. This is also supported by the medians of the KL divergence scores where the median is $0.10$ for \name{} and $0.17$ for \baseline{}.

\textbf{Tasks 5 \& 6.} These tasks were about finding specific images. All the participants of our study were successful in finding the correct images using both the \baseline{} and \name{}.

\textbf{Task 7.} The participants were asked to find labeling errors from misclassified images. Unlike Tasks 5 and 6, multiple correct answers exist. We assessed the images selected by our participants and divided them into three categories: \textit{reasonable, somewhat reasonable, not reasonable}. Based on our assessment of 20 images found 
by 20 participants,
with \baseline{}, 12 are reasonable and 3 are somewhat reasonable; with \name{}, 15 are reasonable and 3 are somewhat reasonable.
This indicates that \name{} is likely more helpful in finding potential anomalies in image datasets.
The images in \name{} are divided into clusters with distinguishable boundaries,
which makes it more convenient to systematically inspect a large number of images 
than with \baseline{}.

\subsubsection{Evaluation of post-questionnaires}
Each participant answered 10 questions in two separate post-questionnaire forms: one for \name{} and one for \baseline{}. 
They provided ratings on a 7-point Likert scale (7 being strongly agree). The questions and their average ratings are shown in \autoref{table:ratings}.

\begin{table}[!tb]
\centering
\begin{tabular}{l @{\hskip3pt}c@{\hskip5pt}  @{\hskip5pt}l@{\hskip3pt} }
\toprule
Question & 
 \baseline{} & \name{} \\\midrule
Easy to learn how to use & \textbf{6.45} & \hskip5pt 6.30\\\hline
Easy to use & 6.00 & \hskip5pt 6.00\\\hline
Helpful for overview & 5.95 & \hskip5pt \textbf{6.45}\textsuperscript{$\circ$}\\\hline
Helpful for detailed analysis & 5.15 & \hskip5pt \textbf{6.05}\textsuperscript{$\ast$}\\\hline
Helpful for finding specific images & 5.10 & \hskip5pt \textbf{5.75}\textsuperscript{$\circ$}\\\hline
Helpful to identify image categories & 5.70 & \hskip5pt \textbf{6.20}\textsuperscript{$\circ$}\\\hline
Helpful to discover new insights & 5.25 & \hskip5pt \textbf{6.00}\textsuperscript{$\circ$}\\\hline
Confident when using the tool & 5.85 & \hskip5pt \textbf{6.05}\\\hline
Enjoyed using the tool & 6.10 & \hskip5pt \textbf{6.40}\\\hline
Would like to use again & 5.80 & \hskip5pt \textbf{6.65}\textsuperscript{$\ast$}\\
\bottomrule
\end{tabular}
\vspace{2pt}
\caption{Participants' average ratings for the two visualizations. \name{} outscored \baseline{} in 8 out of 10 questions. Bold indicates higher average ratings. $\ast$ and $\circ$  indicate 95\% and 90\% statistical significance in one-sided Wilcoxon signed-rank tests, respectively.
}
\vspace{-2pt}
\label{table:ratings}
\end{table}

The results indicate that 
\name{} received higher ratings than \baseline{} in 8 out of 10 questions. 
The \baseline{} received a better rating for only the first question regarding the learnability of the visualization.
This is reasonable as \baseline{} supports fewer interactions than \name{}. 
From the ratings of several important aspects of image visualizations, \name{} is found to be statistically significantly more preferable than \baseline{}, such as getting an overview, performing detailed analysis, identifying image categories, and discovering new insights. 
Moreover,
participants on average inclined more towards \name{} than \baseline{} in mentioning their eagerness to use the tool again. 

\subsection{Discussion}

We observed participants' usage while they performed the tasks.
Based on their usage patterns, we have made a few important findings.

\textbf{\name{} provides a more structured workflow.} Compared to \baseline{}, it is easier to assess or follow how a user makes certain decisions with \name{}. In \name{}, the presence of clusters and the hierarchical relationships within them provide significant semantic information to the participants when they create groups or search images based on certain properties. One participant said: 
\emph{``The clustering of \name{} was very intuitive, more so than the grid one where the boundaries between groups were not clearly defined. The ability to click into different levels of clusters was very useful as well.''}

\textbf{\name{} helps with extracting more specific properties.} Using the semantic information provided by \name{}, the participants could find more detailed information about different image groups. This is more evident with Task 3 where the participants worked with the images of ships and dogs to find two large groups that have high classification accuracy and specific properties. With \name{}, the participants mentioned more specific properties compared to \baseline{}. For example, regarding dogs, \name{} users described their eyes, hair length, and facial structure in addition to generic properties such as size, color, and background. With the \baseline{}, participants mostly described  groups using only generic properties.

\textbf{Image search can be narrowed down more with \name{}.} The hierarchical relationships within the clusters helped the participants narrow their search for a particular image. 
With \name{}, they easily found specific clusters with more images similar to the one they were looking for. The sub-clusters present within a cluster then helped them further narrow the search space. On the other hand, with \baseline{}, they had to check a large group of images as there is no structured way of narrowing the search. One participant said:
\emph{``With the treemap, the ability to narrow down the search without having to recompute the grid size every time, having some predetermined way of organizing the images, and having the images broken up into clusters made it very easy to scan through the images without getting lost. I was able to quickly filter the exact things I was looking for.''}

\textbf{Cluster summary provided with \name{} is helpful.} 
\name{} provides information about each cluster and sub-cluster, such as the number of images and classification accuracy.
The participants found this information useful, especially for Tasks 3 and 4. One participant expressed their liking by saying:
\emph{``I like the clusters having details like how many images and the accuracy. Also, the outline of the different clusters having different sizes helped.''}

\subsection{Limitations}
\label{sec:study-limitations}
No empirical study is perfect. We discuss threats to validity.

\textbf{Different mechanisms for exploration.} While \name{} users can navigate tree-structured data at multiple levels, \baseline{}
does not create a hierarchy by default. 
These differences make \name{} not-surprisingly do better with deeper levels of hierarchical analysis. 
Our intent was to compare our method against a popular baseline for ML practitioners, our target population.
No matter our intent, the threat that any hierarchical method might show similar improvement over the baseline \baseline{} should still be considered.

\textbf{Types of images shown.}
An important potential threat to validity comes down to the image data we used. Depending on the background of the participants, other factors may explain differences in results, such as familiarity of images. 
In addition, the types of images are potentially an ecological threat to validity. In the real world, datasets may contain more diverse, complicated, and noisier images than what is contained in the CIFAR datasets used in our study.
For the purposes of the study, it was necessary to limit the scope for reasonable comparison.

\section{Experiments: Distance Preservation}
\label{sec:numerical}

Lastly, we evaluate the quality of the cluster structures generated from \name{} computationally.
We quantitatively measure \textit{$k$-nearest neighbor accuracy}--how well \name{} preserves the top-$k$ nearest neighbors in the original high-dimensional space.

\subsection{Setup}
We measure the number of common images in the top-$k$ nearest images between one of the techniques and the original high-dimensional representation of data, while varying $k$ (i.e., the size of nearest neighbor list). 
This is a common way to evaluate the quality of DR methods~\cite{wang2020understanding}.
The techniques we compare are:
(1) t-SNE, (2) \baseline{} (described in \autoref{sec:tsnegrid}), and (3) \name{}.
We performed this experiment over 12 different datasets:
CIFAR-10,
CIFAR-100, and
10 subsets of CIFAR-10, each from one of the 10 classes.
All are trained with ResNet50 (same setup described in \autoref{sec:datasets-and-models}), but for the first two, the high-dimensional representations were taken from the last hidden layer, while those for the 10 subsets were taken from the second-to-last hidden layer.

While we compute Euclidean distances between 2-D points for ranking similar images in t-SNE and \baseline{} which assigns a ($x$, $y$) value to each data point,
\name{} needed a different methodology
because it additionally encodes hierarchical structures using treemaps.
We define a distance between two images $\textbf{x}_i$ and $\textbf{x}_j$ in \name{} by
measuring the distance from the node for $\textbf{x}_i$ in the dendrogram tree to the nearest common ancestor node between $\textbf{x}_i$ and $\textbf{x}_j$. This can be thought of as how many times a user needs to zoom-out from the leaf node for $\textbf{x}_i$ to reach to the cluster where both $\textbf{x}_i$ and $\textbf{x}_j$ belong to.

\begin{figure}[!bt]
    \centering
    \includegraphics[width=\columnwidth]{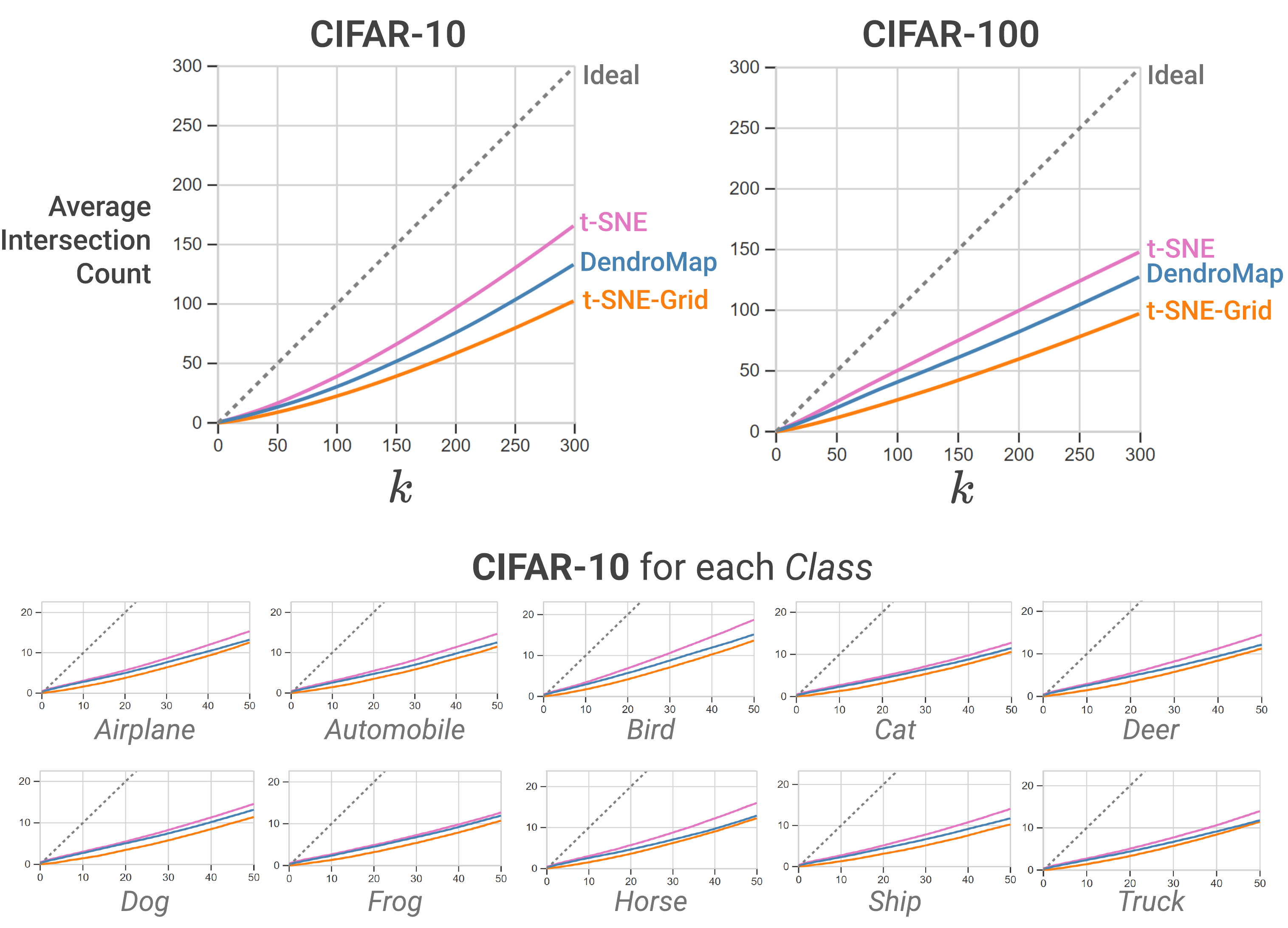}
    \vspace{-16pt}
    \caption{Average for the number of common $k$-nearest neighbours between t-SNE, \baseline{}, or \name{} and high-dimensional representations of images. 
    For all 12 datasets we tested, \name{} preserves the top-$k$ images better than \baseline{}. 
    }
    \label{fig:numerical}
\end{figure}

\subsection{Results}
\label{sec:numerical-result}

Figure~\ref{fig:numerical} shows the results.
For each of the 12 plots,
the $x$-axis represents $k$ (in $k$-nearest neighbor) and the y-axis represents the average number of common images in two top-$k$ image lists.
We display up to 300 for 10,000 image datasets and 50 for the class-level CIFAR-10 datasets
As shown in the figure, in all cases,
t-SNE outperforms the other two, as we can expect, because t-SNE is designed to optimize this metric.
When comparing \name{} and \baseline{}, \name{} shares more top-$k$ nearest neighbors with the high-dimensional representations than \baseline{} for all 12 datasets.
This indicates that \name{} preserves the local similarity structures better than \baseline{}.

\section{Limitations and Future Work}

\indent\indent
\textbf{Computational scalability of clustering.}
The agglomerative clustering algorithms can be a bottleneck when scaling \name{} to larger datasets.
The na\"ive algorithms grow by $O(n^3)$ in time but can be brought down to $O(n^2)$ with optimizations~\cite{mullner2011modern}. The t-SNE method runs with a time complexity $O(n^2)$ and can use approximation to get to $O(n \log n)$~\cite{van2014accelerating}. Although clustering is less efficient, it only needs to be computed once for interactive use in \name{}. 
For the CIFAR-10 test set with $10,000$ images,
the clustering algorithm took 36.8 seconds compared to 32.0 seconds for t-SNE\footnote{Agglomerative clustering was computed with Ward linkage using \href{https://docs.scipy.org/doc/scipy/reference/generated/scipy.cluster.hierarchy.linkage.html}{SciPy}, and t-SNE was computed with the default parameters using \href{https://scikit-learn.org/stable/modules/generated/sklearn.manifold.TSNE.html}{scikit-learn}.} (ran on macOS 12.4, 2.6 GHz 6-Core Intel Core i7 cpu). 
Future work can investigate more efficient strategies to create hierarchical structures of data.

\textbf{Comparison with other tree construction methods.} In the user study, we compared \name{} with t-SNE, the most well-known technique (specifically \baseline{});
however, as noted in the limitations in \autoref{sec:study-limitations}, t-SNE does not create an explicitly hierarchical structure. In the future, \name{} can be compared against a variety of other  techniques (e.g., H-SNE~\cite{pezzotti2016hierarchical}) to evaluate the effectiveness of algorithms that produce hierarchical structures.

\textbf{Interactive refinement of tree structures.}
While the agglomerative clustering algorithms generate hierarchical structures that allow users to flexibly specify the number of clusters to be displayed, 
the formed structures may not be ideal for some cases. 
Visualization researchers have extensively studied interaction methods for steering and refining clustering results~\cite{yang2020interactive,choo2013utopian}.
Future research challenges include designing user interactions for refining clustering results in \name{}.

\textbf{Using interpretable attributes for tree construction.}
We used embedding vectors extracted from deep learning models as input to clustering algorithms, but
alternative methods may help people better interpret substructures of each cluster in \name{}.
For example, representing each image with human-understandable concepts~\cite{kim2018interpretability,zhao2021human} or additional resources~\cite{xie2018semantic} may make each dimension more interpretable.
Alternatively, integrating information about each dimension of the embedding vectors into the interface using explainable AI methods can also be helpful~\cite{olah2018the,hohman2019summit}.

\textbf{Formalizing interaction operations.}
Several data manipulation operations can also be provided in \name{}.
For example, sorting images within each node by user-specified criteria (e.g., prediction scores) 
or splitting and zooming into only a subset of nodes~\cite{bisson2012improving,yang2020interactive}.
Formalizing these types of operations would allow for more flexible user exploration.
Integrating some ideas presented in the unit visualization literature~\cite{park2017atom,wexler2019if,ren2016squares}, such as horizontally or vertically separating space based on categorical attributes in Facets-Dive~\cite{wexler2019if,facets}, into the treemap context would also be an interesting future direction.

\acknowledgments{
We thank Eric Slyman for their feedback.
This work was supported in part by Google Cloud (GCP19980904), NAVER AI Lab, NSF and USDA NIFA (2021-67021-35344), and DARPA (N66001-17-2-4030).
}

\bibliographystyle{abbrv-doi-hyperref}

\bibliography{references}
\end{document}